\documentstyle[preprint,aps]{revtex}
\begin{document}
\draft
\title{
Single particle relaxation in a random magnetic field.
}
\author{E. Altshuler$\sp{1,2}$, A.G. Aronov$\sp{1,2,\dagger}$,
A.D. Mirlin$\sp{1\ddagger}$ and  P. W\"{o}lfle$\sp{1}$}
\address{
$\sp{1}$ Institute f\"{u}r Theorie der Kondersierten Materie,
  Universit\"{a}t Karlsruhe, 76128, Karlsruhe, Germany}
\address{ $\sp{2}$ Department of Condensed Matter Physics,
 The Weizmann Institute of Science, Rehovot, 76100, Israel
}
\date{March 2, 1994}
\maketitle
%\narrowtext
\begin{abstract}
We consider the single particle relaxation rate of 2D
electrons subject to a
random  magnetic field.
The density of states (DOS) oscillations in a uniform
magnetic field (in addition to a random one)
are studied.
It is shown that the infrared singularities
appearing in perturbation theory for the
single particle relaxation rate are spurious
and can be regularized by a  proper  choice of gauge.
We define a gauge invariant single particle relaxation time
as a parameter
describing the amplitude of DOS oscillations.
Unlike the conventional case of random potential
scattering the broadening of the Landau levels
does not depend on magnetic field.
\end{abstract}
\narrowtext
The problem of a charged quantum particle moving in a
2D static random magnetic field has attracted
considerable interest recently.
Models of this type are supposed to describe
effectively a number of important physical systems,
such as a state with spin-charge separation in high-$T\sb{c}$
superconductors \cite{il,nl,aw} and a quantum Hall system near the
filling factor $\nu =1/2$ \cite{kz,hlr}.
In addition, the direct experimental realization of static random
magnetic field has been reported recently \cite{geim}.

In physical realizations of such a system
one is interested in a random {\it magnetic field}
distribution characterized by some correlation length.
This is unlike the usual case of a disordered system,
where the {\it electric potential}
is a random variable with finite range correlations.
Here the corresponding quantity is the vector potential,
which will be governed by a spatially long-ranged correlation
function, even for short-ranged magnetic field correlations.
Consequently, similar to the case of long-range potential
fluctuations, small angle scattering will play an important
role, and might even cause the total scattering cross-section
to diverge.
On the other hand, the transport scattering cross-section,
which is weighted with the momentum transfer squared,
is expected to be well-behaved.

In a previous paper \cite{amw} the transport
properties of the model were studied.
It was shown there that although the single
particle relaxation rate $1/\tau$ is plagued by
infrared divergence problems,
the electrical conductivity is determined by the transport
(momentum) relaxation
time $\tau\sb{tr}$, which can be  calculated perturbatively
(see also \cite{km}) .
As far as the critical behavior of the model
is concerned, the problem could be mapped onto a nonlinear
supersymmetric $\sigma$-model of interacting matrices
with unitary symmetry.
The diffusion constant
(or equivalently the dimensionless conductance) $g\sb{0}$
was shown to be determined by $\tau\sb{tr}$
rather than $\tau$.
This means that all the states are localized within
the localization length
$\xi\propto \exp{(\pi g\sb{0})\sp{2}}$.
On  shorter length scales the transport is governed by the
usual diffusion law, in  contrast to the suggestion made
in \cite{weat} that the particle diffuses according to a
logarithmic diffusive law.
We believe this to be an artefact of an  improper
approximation employed in \cite{weat},
as will be discussed in the end of the paper.

In the present paper we consider the single particle
properties of the model.
There are two related problems that have to be dealt with:
the existence of infrared divergences in perturbation theory
and the gauge dependence of the Green's function.
The latter has been studied recently \cite{ai} in a path
integral formulation.
In order to circumvent the gauge problem
the authors of \cite{ai}
proposed to introduce a gauge invariant Green's function,
obtained by multiplying the usual single particle Green's function
by a phase factor.
The phase was chosen as the one caused by integrating the vector potential
along a  straight line connecting the end points.
The Green's function defined in this way was evaluated in
the quasiclassical approximation.
Obviously, this quantity depends on the choice of the integration path
determining the phase.
The physical meaning and possible applications of this Green's function
are far from clear.
The same object later was studied in \cite{km},
and the results of \cite{ai} were reproduced.

We define below the single particle relaxation time as the
parameter of the broadening of the Landau levels
in a weak, constant magnetic field (in the presence of the
random one). This is an unambiguous experimentally meaningful
parameter which is definitely gauge invariant.
It can be determined by measuring the de Haas -- van Alphen
or Shubnikov--de Haas oscillations.
Note that the constant magnetic field appears naturally and inevitably
for the direct experimental realization of the model \cite{geim}.

In a previous paper \cite{amw} it was observed that the  calculation
of the single particle relaxation time is complicated by
the infrared divergence of the corresponding diagrams.
Meeting analogous problems in the slave-boson representation
of the $t-j$ model the authors of \cite{nl}
concluded that the two-point auxiliary particle
Green's function should vanish identically.
These authors concentrated on  the consideration of the
transport time.
We will show here that the infrared divergence is fictitious
and can be removed by a gauge transformation.
The procedure we use for this
purpose is somewhat analogous to that used in \cite{ef,raikh}
for separation of the low-momentum
transfer scattering processes in a random potential with
long-ranged correlations.
Although the resulting Green's function has no infrared singularity
it still changes under gauge transformations.
This should not pose a problem for the one point Green's function,
which is gauge invariant and the imaginary part of which determines
the  density of states (DOS).

We show that  the use of the
self-consistent Born approximation \cite{ando,raikh,we}  for
calculation of the self-energy violates the gauge invariance
and leads to an ambiguous result for $\tau$.
The ambiguity can be resolved, however,  by using the formalism
of path integrals \cite{feinm} in coordinate space.
We obtain a Gaussian, rather than Lorentzian, shape of the
broadening of the Landau levels which explains the difficulties
arising when one tries to treat the problem by means of the
conventional diagram technique.

Consider a charged spinless particle
(with the mass $m$ and charge $e$) in a 2D layer,
in a random magnetic field $\bbox{ H}(\bbox{r})$ perpendicular to the layer.
The random magnetic field is supposed to be Gaussian distributed with
a zero mean value and correlator:
\begin{equation}
\langle H(\bbox{r})H(\bbox{r}\sp{\prime})\rangle
= \langle h\sp{2}\rangle\delta(\bbox{r}-\bbox{r}\sp{\prime})
\label{hcor}
\end{equation}
The vector potential correlator, in  momentum space,
can be taken in the form:
\begin{equation}
\langle A\sb{\alpha}(\bbox{q})A\sb{\beta}(-\bbox{q})\rangle
=\delta\sb{\alpha\beta} {\langle h\sp{2}\rangle\over q\sp{2}}
=D(\bbox{q})\delta\sb{\alpha\beta}
\label{acor}
\end{equation}
Applying the standard perturbation theory with the
correlation function (\ref{acor}), one encounters divergences
generated by the singular nature of (\ref{acor}) in the limit
$q\rightarrow 0$.
A similar problem was considered in \cite{ef,raikh}
for the case of the long-ranged Coulomb potential.
Exploiting the ideas from \cite{ef,raikh} we represent the correlator
$D(q)$ in the form:
\begin{eqnarray}
&&D(\bbox{q})=D\sb{0}(\bbox{q}) + \tilde{D}(\bbox{q}) \nonumber \\
&&D\sb{0}(\bbox{q})=d\sb{0}\delta (\bbox{q});\ \ \
d\sb{0}=\int\sb{q<Q}d\sp{2}\bbox{ q}\: D(\bbox{q})
\label{corsep}
\end{eqnarray}
where $Q$ is an arbitrary momentum cut-off, which serves to separate
low- and high- momentum contributions.
It can be shown that the averaged single-particle Green's function
%\begin{equation}
$
G(E,\bbox{ p})=
\left<\left[E-{1\over 2m}(\bbox{ p}
-{e\over c}\bbox{ A})\sp{2} \right]\sp{-1} \right>
$
%\label{agf}
%\end{equation}
can be rewritten in the form :
\begin{equation}
G(E,\bbox{p})={1\over 2\pi d\sb{0}}
\int d\sp{2}\bbox{ A}\: e\sp{-{\bbox{ A}\sp{2}/2d\sb{0}}}\,
\tilde{G}\left(E,(\bbox{ p}-{e\over c}\bbox{ A})\right)
\label{ggf}
\end{equation}
where $\tilde{G}$ differs from $G$ in that the correlator
$\tilde{D}(q)$ (rather than $D(q)$) is used for the
averaging over the disorder.

The crucial difference between (\ref{ggf}) and the analogous
expression in \cite{ef,raikh} is that on the r.h.s. of (\ref{ggf})
the averaging enters by the shift of the momentum $\bbox{ p}$,
whereas in  \cite{ef,raikh} it entered as a shift of the chemical
potential.
In principle, the transformation (\ref{ggf})
affects the Green's function in a non-trivial way.
However, if we consider the one-point Green's function
$\int d\sp{2}\bbox{ p}\: G(E,\bbox{ p})$  \cite{note},
shifting the integration variable
$\bbox{p}$ we obtain:
\begin{equation}
\int d\sp{2}\bbox{ p}\: G(E,\bbox{ p}) =
\int d\sp{2}\bbox{ p}\:\tilde{G}(E,\bbox{ p})
\label{ginv}
\end{equation}
This is nothing else than the manifestation of gauge invariance.
The correlator $D\sb{0}(q)$ corresponds to a spatially uniform vector
potential, {\it i.e.}  to zero magnetic field.
Thus, one can use $ \tilde{D}(q) $ instead of $D(q)$ for the calculation,
since it does not affect any gauge invariant quantity.
The advantage of this simple trick is that now all the infrared
divergences are removed from the diagrams.
This can be checked for any particular diagram.
Demonstrate the cancelation of the infrared
divergences for the sum of non--crossing diagrams, which
represents SCBA. For simplicity we consider the case
when the constant magnetic field is equal zero.
We have for the imaginary part of the self-energy:
\begin{eqnarray}
{\hbar /2\tau( E ,\bbox{p})} & = & v\sb{0}\sp{2}
\int\sb{q<Q} {d\sp{2}\bbox{q}\over q\sp{2}}
  \left\{
{{(2\bbox{ p} + \bbox{ q})\sp{2} \hbar /2\tau\sp{\prime}}\over \left(  E -
{(\bbox{ p} + \bbox{ q})\sp{2}/2m}\right)\sp{2} +
\left({\hbar /2\tau\sp{\prime}}\right)\sp{2}}
 \right.  \left. -
{{(2\bbox{p})\sp{2}\hbar /2\tau}\over \left(  E -
{\bbox{ p}\sp{2}/2m}\right)\sp{2} +
\left({\hbar /2\tau}\right)\sp{2}} \right\} \nonumber \\
& + &
v\sb{0}\sp{2}\int\sb{q>Q} {d\sp{2}\bbox{q}\over q\sp{2}}\:
{{(2\bbox{ p} + \bbox{ q})\sp{2}\hbar /2\tau\sp{\prime}}\over \left(  E -
{(\bbox{ p} + \bbox{ q})\sp{2}/2m}\right)\sp{2} +
\left({\hbar /2\tau\sp{\prime}}\right)\sp{2}}
\label{cutoff}
\end{eqnarray}
where $v\sb{0}\sp{2}=e\sp{2}\langle h\sp{2}\rangle/4m\sp{2}c\sp{2}$ is the
characteristic
velocity of the problem, $\tau=\tau( E ,\bbox{ p}) $,
$\tau\sp{\prime}=\tau( E ,\bbox{ p}+\bbox{ q}) $.
It is easy to see that the first integral does not diverge at small
$\bbox{q}$ anymore.
Since we are interested in the mass-shell value of $\tau$,
we take $ E =\bbox{ p}\sp{2}/2m$ and suppose that
$ E\tau\gg \hbar$.
The integrals are dominated by a vicinity of the mass-shell,
where one can take $\tau\sp{\prime}=\tau$.
A straightforward calculation gives :
\begin{equation}
\left({\hbar\over 2\tau}\right)\sp{2}
={2\over\pi}Emv\sb{0}\sp{2}
\ln{\left( {\hbar m\over Q\sp{2}\tau}\right)}
\label{tscba}
\end{equation}
As was discussed above, due to the gauge symmetry,
no  observable  quantity can depend on
the cut--off parameter $Q$.
However, repeating the same derivation for the case of
non--zero constant magnetic field, one obtains the same
ambivalent logarithm for the amplitude of oscillations
of the density of states,
which is definitely a gauge invariant quantity.
The reason for  obtaining such an ill-behaved
single particle time is that we implicitly
expected the Green's function to be of the usual,
Lorentzian, form:
\begin{equation}
G(E,\bbox{ p})=
\left(  E-\bbox{ p}\sp{2}/2m +i\hbar /2\tau\right)\sp{-1}
\label{lor}
\end{equation}
Actually one should not expect this form of the
Green's function.
Assume, for a moment, such a form for a particular
value of $Q$.
Then any change of $Q$ transforms the Green's function
in a way described by equation (\ref{ggf}),
and the Green's function acquires a Gaussian form
rather than the Lorentzian one.
Thus the assumption that the Green's function can be
described by (\ref{lor}) contradicts the gauge invariance,
and leads to the above mentioned ill behavior of the single
particle time.

The physical reason for the appearance of the
non--gauge--invariant logarithm in (\ref{tscba}) is the following.
Consider the open trajectories for the two--point Green's function
in a path integral representation.
Gauge freedom means that we can add arbitrary
far remote magnetic
flux tubes to the system which will give rise
to  additional random phases.
The choice of $Q$ specifies the scale for the influence of
the remote random fluxes on the Green's function.
This  ambiguity is removed if we restrict the consideration
only to closed trajectories ({\it i.e.} one--point Green's function)
 on which the random phase is a well
defined quantity determined by the random flux enclosed by
the trajectory.
However, in order to calculate the self-consistent single particle
relaxation rate we are forced to consider open
trajectories, which, having gauge ambiguous random phases,
cause the appearance of the
ambiguous logarithm in the corresponding relaxation
time (\ref{tscba}).

To overcome these difficulties, arising in the diagrammatic
%due to the SCBA
approximation,
let us consider a different approach to the problem.
In the following we use the quasiclassical approximation
{\it i.e.} assume
$E\tau\gg \hbar$ and $E/\omega\gg \hbar$,
where $E$ is the Fermi energy,
$\omega=eH\sb{c}/mc$ is the cyclotron frequency,
$H\sb{c}$ is a uniform magnetic field.

It is known \cite{feinm} that the one-point Green's function
can be obtained, in the quasiclassical limit,
by calculating the path integral
over the closed classical trajectories.
In our case, after averaging over the random
magnetic field, the action on such trajectories
is \cite{weat}:
\begin{equation}
i{\cal S}=i{e\over c}H\sb{c}s\sb{o}-
\left({e\over c}\right)\sp{2}{\langle h\sp{2}\rangle\over 2} s\sb{no}(E)
\label{action}
\end{equation}
Here $s\sb{o}$ and $s\sb{no}$ denote the oriented
and non--oriented (Amp\`{e}rean) areas of the trajectory respectively.
The first term describes the phase acquired by the particle
due to the flux of the constant magnetic field and the
second appears due to averaging over the random one.
For  trajectories  without self-intersections the areas can differ only
in sign ($s\sb{no}=\vert s\sb{o} \vert$).
The non--oriented area of a self-intersecting trajectory depends
non--trivially  on its topology,
{\it e.g.} for the closed trajectory with a winding number
$k$ the non--oriented area is proportional to $k\sp{2}$.

We will consider the classical trajectories
determined by the first term in (\ref{action}),
and treat the second term as a perturbation.
The criterion of the applicability of this approximation
will be discussed later.
Classical trajectories in a magnetic field are circles
with a cyclotron radius $R\sb{c}=p/(m\omega)$.
Calculating  the action on the classical trajectory
${\cal S}(E)=(e/c)H\sb{c}\pi R\sb{c}\sp{2}=2\pi E/\omega$,
we can write the Green's function as a sum over classical
trajectories with different winding numbers \cite{hake}:
\begin{equation}
G(E)={-im\over 2 \hbar \sp 2}\left[1+2\sum\sb{k=1}\sp{\infty}
\exp { \left\{ 2\pi k i\left[ {E\over\hbar \omega} +{1\over 2}\right]
- 4\pi k\sp{2}{E\over\hbar \omega\sp{2}}mv\sb{0}\sp{2}
\right\} }\right]
\label{odos}
\end{equation}
Comparing the two terms in the exponent in (\ref{odos})
one can see that the perturbative treatment of the
random field is valid, provided
$E\gg\omega\gg mv\sb{0}\sp{2}$.
To obtain explicitly the shape of the broadened
Landau level we use the Poisson formula to
resum (\ref{odos}) and represent
the DOS $\rho(E)=-{1\over\pi}$ Im$G(E)$ as a sum
over the energy quantum numbers  $N$:
\begin{equation}
\rho(E)={1\over 2\pi l\sb{H}\sp 2}\sum\sb{N=0}\sp{\infty}
\sqrt{1\over\pi}{\tau(E)\over \hbar}
\exp{\left\{ -\left[E-\hbar \omega(N +{1\over 2})\right]\sp{2}
{\tau\sp{2}(E)\over\hbar\sp{2}}
\right\} }
\label{ldos}
\end{equation}
where $l\sb{H}=\sqrt{c\hbar/eH\sb{c}}$ is the magnetic length.
This formula evidently shows that the Landau levels acquire
the Gaussian form. This fact explains the difficulties which
one meets trying to use the conventional diagram technique,
which  implicitly assumes a Lorentzian broadening of
the Landau levels.
The Gaussian shape of the Landau levels appears here as a result
of the quasiclassical approximation.
For the case of the potential scattering
one can carry out the analogous derivation, when the potential is
smooth enough to use the quasiclassical approximation.
This also leads to a Gaussian shape of the DOS in agreement
with \cite{raikh}.

The characteristic width of the levels is :
\begin{equation}
 {\hbar\over 2\tau(E)}=\left( {1\over\pi}Emv\sb{0}\sp{2}\right)\sp{1/2}
\label{tau}
\end{equation}
Note that this expression surprisingly coincides
with one obtained by SCBA, except for
the ambivalent logarithm.
The cyclotron frequency does not enter the expression for
$\tau$, thus one can consider $\tau$ as a
meaningful parameter even in the limit of
zero magnetic field $H\sb{c}$.

Equation (\ref{odos}) describes DOS
as a function of energy.
For the case of weak uniform magnetic field ($\omega\tau\ll 1$),
omitting all the  harmonics except
the first one
we obtain for the oscillating DOS $\rho\sp{osc}(E)$:
\begin{equation}
\rho\sp{osc}(E)=-{m\over \pi \hbar \sp 2}
\exp{\left\{-{\pi^2\over [\omega\tau(E) ]\sp{2}}\right\} }
\cos{\left(2\pi {E\over\hbar \omega}\right)}
\label{1dos}
\end{equation}
One can see that the amplitude of the  DOS  oscillations
is proportional to
$\exp{\left\{ -\pi^2/(\omega\tau)\sp{2}\right\} }$,
in contrast to the typical case of random potential
scattering when the amplitude of oscillations is proportional to
 $\exp{\left\{ -\pi/(\omega\tau)\right\} }$.

For the case of strong magnetic field  ($\omega\tau\gg 1$)
we obtain the DOS as a sequence of Gaussian
broadened Landau levels with a width given by (\ref{tau}).
Note that the width does not increase with the magnetic
field, like it does  for the usual case of the potential scattering
\cite{ando,raikh,we}.
The reason for this unusual behavior is that the random phase
acquired by the particle moving along its classical
trajectory is proportional to the area enclosed by the
trajectory
rather than the length of the trajectory.

To make this point clear consider the difference between our
problem and that of the DOS oscillations
in the uniform  magnetic field in the presence of a smooth
random potential $V(\bbox{r})$ with correlator
$\langle V(\bbox{r})V(\bbox{r}\sp{\prime})\rangle
=\xi\sp{-2}U
\exp\left\{- (\bbox{r}-\bbox{r}\sp{\prime})\sp{2}/\xi\sp{2} \right\} $,
where $\xi\gg l\sb{H}$ is the correlation length.
For our case the mean square random phase gained by the particle on
its classical trajectory with a winding number $k$ is
$$\langle\phi\sp{2}\rangle\sim\left({e\over c}\right)\sp{2}\langle
h\sp{2}\rangle s\sb{no}\sim \left({e\over c}\right)\sp{2}\langle
h\sp{2}\rangle k\sp{2}R\sb{c}\sp{2}\sim Emv\sb{0}\sp{2}t\sp{2}$$
where $t=2\pi k/\omega$ is the time spent by the particle
on the trajectory.
In case of potential scattering the analogous estimate
gives instead:
$$\langle\phi\sp{2}\rangle\sim {U\over v\sb{F}\sp{2}\xi}R\sb{c}k\sp{2}
\sim {U\over  \xi R\sb{c}}t\sp{2}$$
where $v\sb{F}$ is the Fermi velocity.
For both cases we can estimate $\tau$ from the condition
$\langle\phi\sp{2}\rangle\sim 1$ which leads in our
case to the result (\ref{tau}) and in the case of potential
scattering reproduces results of \cite{raikh}.

Proportionality of $\langle\phi\sp{2}\rangle$ to the area,
rather than to the length, of the trajectory
actually determines the peculiarity of the
random magnetic field scattering.
Until the geometry of a certain
problem is specified, any statement about
single particle relaxation time will be senseless.
We considered the DOS oscillations
in a constant magnetic field, thus
the characteristic area for this
specific problem is the non-oriented area of the cyclotron circle:
$s\sb{no}=(v\sb{F}t)\sp{2}/4\pi$.
If one would consider, say, the double-slit experiment \cite{feinm}
the area which enters the problem will be determined by the
characteristic length of the trajectories $L$ and the characteristic
angle $\alpha$ between them:  $s=L\sp{2}\sin\alpha$.
Thus one can estimate  the single particle relaxation time in this
case as follows:
$1/\tau\sp{2}\sim Emv\sb{0}\sp{2}\sin\alpha$.
Appearance of $\sin\alpha$ in this expression reflects the dependence
of $1/\tau$ on the geometry of the problem.

Though the object proposed in \cite{ai}
(and later in \cite{km}) has, of course, the most
evident choice of trajectory area for the two-point
single particle Green's function, it is hard to
imagine the physical problem in which it could enter.
The Ref.\cite{weat} addressed the question of the transport
in the random magnetic field in terms of the path integral
formalism.
In this case, the relevant trajectories are of two-particle
diffuson nature.
Thus the the typical single particle trajectories, and their areas,
considered in \cite{weat} are of dubious relevance to the transport
problem.

In conclusion, we discussed the single particle relaxation of
2D electrons in a  spatially random magnetic field
(in the presence of an additional uniform field).
It was shown that any attempt to introduce a single particle
relaxation time as a parameter of the Green's function
of the type (\ref{lor}) leads to an ambiguity since it
violates the gauge invariance.
The physically meaningful single particle relaxation time
appears as a characteristic broadening of the Landau levels
(\ref{tau}).
Broadened Landau levels have the Gaussian shape (\ref{ldos}).
The width of the peaks does not increase with the  magnetic field.
The reason for this behavior is that the random
phase acquired by the electron on a classical
trajectory  is proportional
to the area enclosed by the trajectory,
rather than to its length.

We are very grateful to L.P. Pitaevskii for numerous
helpful comments.
A.G.A. appreciates discussions with U.Sivan.
A.D.M. gratefully acknowledges  discussions with
O. Bohigas and E. Bogomolny.
This work was supported by German-Israel Foundation for Research
[GIF, Jerusalem, Israel] (E.A.), Landau-Weizmann program and
Heraeus-Stiftung (A.G.A.), the Alexander von Humboldt Foundation
(A.D.M.)
and by SFB 195 der Deutschen Forschungsgemeinschaft (P.W.).


\begin{references}
\bibitem{note}The same is true for the diagrammatic
representation of any gauge invariant quantity,
where the integration over all the electrons momenta is
performed.
\end{references}
\end{document}